
\font\titlefont = cmr10 scaled \magstep2
\magnification=\magstep1
\vsize=22truecm
\voffset=1.75truecm
\hsize=15truecm
\hoffset=0.95truecm
\baselineskip=20pt

\settabs 18 \columns

\def\b{\bigskip}
\def\bb{\bigskip\bigskip}

\def\ce{\centerline}

\def\no{\noindent}

\rightline{ISJ-5268}
\rightline{December 1993}
\bb

\ce{\titlefont {Electroweak Sphaleron in Large Higgs Boson Mass Limit}}

\b

\ce{Xinmin Zhang and Bing-Lin Young}
\b
\ce{Department of Physics \& Astronomy \& Ames Laboratory}
\ce{ Iowa State University,}
\ce{ Ames, Iowa
 50011}

\b
\bb
\ce{\bf ABSTRACT}
\b
Since the triviality argument of the Higgs sector requires the existence of
new physics beyond the standard model, there should exist a cutoff
$\Lambda$ beyond which the standard model will breakdown.  The cutoff
can be determined from the position of the Landau pole.
We study the effect of this cutoff on the energy of the electroweak
sphaleron, $E^{spha}$. We found that
$E^{spha}$ becomes arbitrarily large as the Higgs boson mass increases,
caused by the existence of a dimension eight operator.
This is in contrast to the well-known result, which is held in the
standard model
and a wide class of its extensions, that $E^{spha}$ is stable
against the variation of the
 Higgs boson mass. The physical meaning of this result is discussed.
\bb

\bb
\filbreak

\no {\bf 1.}~~The sphaleron in the electroweak theory plays an essential role
in the calculation of the baryon number violation in colliders and
 at high temperature, because it sets the energy scale at which the nontrivial
structure of the finite energy
 configuration space becomes apparent. More concretely, the
sphaleron gives the
minimal
 height of the energy barrier between the topologically
inequivalent neighbouring
 vacua and this barrier determines the rate
of the (B+L), the baryon number plus lepton number, violating reactions.
The energy of the sphaleron in the standard model (SM), calculated first by
Manton and Klinkhamer in a spontaneously broken SU(2) theory
[1], then improved in [2], is given by $E^{spha} = {2 M_W \over \alpha_W}
         B( \lambda /  g_W^2 )$, where
$M_W$ is the W-boson mass,
$\alpha_W = {g^2_W / 4 \pi} $ and
$g_W$ is the SU(2) coupling constant; $B$ depends on the ratio of the Higgs
quartic coupling, $\lambda$, and
$g_W$. Numerical evaluation gives
$1.5 \leq B \leq 2.7$ when the ${\it bare}$ Higgs boson mass,
$m_H^2 = 2 \lambda v^2$, varies from zero to infinity[F.1].
 Hence the value of $E^{spha}$ is
extremely stable against the change in the Higgs boson mass. Furthermore, the
stability of the sphaleron energy
against the changes
of models for a wide class of models has been demonstrated.  The models
considered include the extensions of the standard
one-doublet Higgs sector and the addition of dimension six operators
to the standard model lagragian[4]. Specifically, the values of $E^{spha}$ in
those extended
models
have been shown to change very little for a given Higgs boson mass,
much less than the variation
of its standard model value due to the change of the Higgs boson mass.

Based on the well-known triviality argument of the Higgs sector of
the standard model, it is often argued that the standard model is an
effective theory and there must be a more fundamental theory beyond SM.
In particular, the Higgs sector of
the standard model will have to be
modified in this case.
The scale of the new physics, parametrized in terms of a cut-off $\Lambda$,
beyond which SM has to be modified, can be estimated by
the location of the Landau pole of the Higgs quartic coupling
constant. Approximately one has[5],
$${
\Lambda ~ \leq v ~exp{( {8 \pi^2 v^2 \over {3 m_H^2} })}
  = v~ exp {({ 2 \pi \over {\alpha_W ( \lambda / g^2_W )} })} ~~~, }
\eqno(1)$$

\no where $v$ is the Higgs vacuum expectation value and
$m_H$ the Higgs boson mass.
 One can see that as Higgs boson mass increases, the cut-off $\Lambda$
decreases and the new physics effects become more important
near the electroweak symmetry breaking scale $v$. In this paper,
we study the sensitivity of the
SM sphaleron energy to new physics required by
the consideration of the ``triviality" of the SM Higgs sector.

The importance of the effect of new physics in the energy regime below
the electroweak energy scale in relation to the cutoff
has been studied by Corteses, Pallante and Petronzio[5] in the context
of the standard model electroweak radiative correction to the LEP
data due to Heavy Higgs.
They concluded that the decreasing of the
location of the Landau pole with increasing Higgs boson mass implies that the
sensitivity
to the effect of the radiative correction to the value of the
Higgs boson mass for heavy Higgs will be
lost due to the sizable cut-off dependence of the theory as $\Lambda$
approaches $v$ for very large $m_{H}$. Generally,
the cut-off effects can be accounted for by adding higher
dimension operators to the standard model lagrangian. In this paper
we demonstrate that the effect of the cut-off can cause
$E^{spha}$ to increase without bound as Higgs boson mass goes to infinity.

\b

\no {\bf 2.} ~~Below we briefly review the SM sphaleron[F.2].
 Using the spherical symmetric ansatz[1],
we can write the static fields in the $W_0 = 0$ gauge,
 $${
W_i^a \sigma^a dx^i = - {2 i\over g_W} f( \xi ) ~dU^{\infty} {( U^{\infty} )}
                ^{-1} ~~~; } \eqno(2.a)$$

$${
\phi = {v \over {\sqrt2}} ~ h( \xi ) ~ U^\infty { \pmatrix{0 \cr
                                1 \cr} } ~~~~~, }\eqno(2.b)$$

\no where
$${
U^\infty = {1 \over r} \pmatrix{ z & x + iy \cr
                              -x + iy & z \cr } ~~~, }$$
\no and $\xi = g_{W} v r$.

The energy functional is given by
$${ \eqalign{ E^{spha} = & \int d^3 x ~ [{1\over 4} F_{i j}^a F^{a i j}
                                        +{| D_i \phi |}^2 \cr
                       & + \lambda {( {|\phi |}^2 - v^2/2 )}^2 ] ~, \cr}
}\eqno(3)$$
\no where
$${
F_{i j}^a = \partial_i W^a_j - \partial_j W^a_i + g_W \epsilon^{abc}
                        W_i^b W_j^c ~; }$$

$${
D_i \phi = \partial_i \phi - {i \over 2} g_W W_i^a \sigma^a \phi
        ~~. }$$

\no The sphaleron energy is obtained by minimizing eq.(3).  This gives
rise to a set of coupled non-linear differential equations involving
$f( \xi )$ and $h( \xi )$. Their solutions determine the sphaleron
energy from (3). The boundary conditions for $f( \xi )$ and
$h( \xi )$ are given by[1]

$${
f( \xi ) \rightarrow \xi^2 ~~{\rm and} ~~h( \xi ) \rightarrow
                  \xi ~~~{\rm for} ~~ \xi \rightarrow 0 ~;}\eqno(4.a)$$

$${
f( \xi ) ~ ~{\rm and} ~~h( \xi ) \rightarrow 1 ~~{\rm for} ~~\xi
    \rightarrow \infty . } \eqno(4.b)$$
\b

To estimate $E^{spha}$, let us use the Ansatz of Klinkhamer and Manton [1],
$${
f( \xi )= {\xi^2 \over {\Xi \over {\Xi + 4 }} }, ~~~{\rm for}~~
                                     \xi \leq \Xi ~~; }\eqno(5.a)$$

$${
f( \xi ) = 1 - {4 \over {\Xi + 4} }
                 exp[{ {1\over 2}( \Xi - \xi ) }] , ~~~{\rm for}~~
                       \xi \geq \Xi ~~, }\eqno(5.b)$$

\no and

$${
h( \xi ) = { \sigma \Omega + 1 \over {\sigma \Omega + 2 } }
           ~{ \xi \over \Omega }, ~~~{\rm for} ~~\xi \leq \Omega ~; }
\eqno(5.c)$$

$${
h( \xi ) = 1 - { \Omega \over { \sigma \Omega + 2 }} {1 \over \xi }
                       exp[ \sigma ( \Omega - \xi )] ,
{}~~~{\rm for} ~~\xi \geq \Omega ~, }\eqno(5.d)$$

\no where $\Xi$ and
$\Omega$ are determined by minimizing the energy functional for a given
values of
$\lambda / g^2_W$. Some of these values given in Klinkhamer and Manton [1]
are listed
in table I to show how $\Xi$ and $\Omega$ varies with
$\lambda / g^2_W$.

\b

\vbox{\tabskip=0pt \offinterlineskip
\def\tablerule{\noalign{\hrule}}
\halign to300pt {\strut#& \vrule#\tabskip=1em plus2em&
  \hfil#& \vrule#& \hfil#\hfil& \vrule#&
    \hfil#& \vrule# \tabskip=0pt \cr \tablerule

&& $\lambda / g^2_W$ && $\Omega $  && $\Xi $
                                                               &\cr \tablerule
&& 0   &&  2.600 &&  2.660   &\cr \tablerule
&& $10^{-3} $ && 2.520 && 2.450 &\cr \tablerule
&& $10^{-2}$  && 2.290 && 2.120 &\cr \tablerule
&& $10^{-1}$ &&   1.900 &&   1.650 &\cr \tablerule
&& 1  && 1.250 && 1.150 &\cr \tablerule
&& 10 && 0.620 && 0.820 &\cr \tablerule
&& $10^2 $ && 0.220 && 0.740 &\cr \tablerule
&& $10^3 $ && 0.070 && 0.730 &\cr \tablerule
&& $\infty$ && 0  && 0.728 &\cr \tablerule \noalign{\smallskip}
& \multispan{7} Table I.
\hfil \cr
&\multispan{7}  \hfil \cr }}

\bb

\no {\bf 3.} ~~Now let's consider the cut-off effects on the SM sphaleron
energy.
We have examined the effect of many higher dimension operators
which involve the Higgs field and are invariant under the standard model
gauge symmetry.  However, we
will concentrate here on the following most interesting operator
  for detail
discussion,

$${
  {\cal O} \sim {1 \over \Lambda^4 }
            {  \{ {(D_\mu \phi)}^\dagger D^\mu \phi \} }^2 ~~~. }\eqno(6)$$

\no The reason why we choose $\cal O$ is that it is the operator
of the lowest dimension that makes $E^{spha}$ diverge in the heavy
Higgs boson mass limit. We present our arguments below.

 Using the spherical symmetric ansatz (2), we have the contribution
of the operator
$\cal O$ to
$E^{spha}$,

$${
\Delta E^{spha} = {g_W \pi v^5 \over \Lambda^4}
                  \int d\xi \{ \xi^2 {( {dh \over {d \xi } })}^4
                        + 4 h^4 {( {dh \over {d \xi} })}^2
 {( 1-f )}^2 + 4 {h^4 \over \xi^2 } {( 1-f )}^4 \}
{}~~. }\eqno(7)$$

\no The last term in the right-handed side causes
 $\Delta E^{spha}$ to diverge
as ${\lambda \over g^2_W} \rightarrow \infty$.
This can be seen clearly by applying
  the ansatz eqs.(5),
although the result is
independent of this particular ansatz.
 The last term of the right-handed side
of eq.(7) dominates for very heavy Higgs,

$${
\Delta E^{spha} \sim \int d \xi {h^4 \over \xi^2 } {( 1-f)}^4
                \rightarrow {1 \over \Omega} ~~. }\eqno(8)$$

\no From table I we see that $\Omega \rightarrow 0$ and
hence $\Delta E^{spha} \rightarrow \infty$
as
$\lambda \rightarrow \infty$.

 The result is not a consequence of the "perturbative" treatment given
above. Incorporating the
operator $\cal O$ directly into the sphaleron differential equations,
and solving the differential equations without using
the particular ansatz (5.a $\sim $ d), our above conclusion still holds.
The reason for the blowup of the sphaleron energy in the limit of
very large Higgs boson mass is the following. The boundary conditions (4.a) and
(4.b) are still valid with the addition of the new term (6) directly
to (3). The corresponding differential equations for $f(\xi)$
and $h(\xi)$ are obtained by minimizing the modified energy expression. In the
limit of large $\lambda$ the
Higgs potential term in (3) requires that ${| \phi |}^2 \rightarrow
  v^2/2,~~ ({\rm equivalently} ~~ h \rightarrow 1 ~ ) $ for $\xi > 0 $ , and
${\phi = 0}$ for ${\xi = 0}$.
This implies that (8) diverges in the limit of very large $\lambda$
or the Higgs boson mass.

\b

\no {\bf 4.} ~~To illustrate the physical meaning of the above result,
we consider a toy model of the dynamical symmetry breaking (DSB) theory.
Let us consider
a one family standard model where there is no elementary Higgs field.
In this model the SM gauge symmetry is broken by the quark condensate
driven by the QCD interaction, where both the electroweak symmetry breaking
scale and the weak gauge boson masses are, of course, very low.
Then the Higgs fields,
$\sigma$ and
the Goldstone pions,
 are composites, made of the up and down quarks.

It is well-known that the strong interaction of a DSB model can produce
various higher dimension operators when the heavy fermions, i.e., the
quarks, are frozen out at energy below the DSB strong interaction scale.
Integrating (freezing) out the heavy fermions will produce $SU(2)$ triplet
and singlet effective operators which are arranged in some gauge invariant way.
Then an operator such as the $\cal O$ naturally exist[6].  Let us write
$\phi = {h \over {\sqrt 2}} \Sigma$ with
$\Sigma$ being the unitary part
of the Goldstone field and $h = H + v$, where
$H$ is the
physical Higgs field.
For the dynamical model,
where no physical Higgs field appears in the effective lagrangian,
 operator $\cal O$ is reduced to
$${
{\tilde {\cal O}} \sim { ( (D_\mu \Sigma)^\dagger D^\mu \Sigma )}^2
 ~~~, } \eqno(9)$$

\no which exists
in the effective lagrangian of pion fields of the QCD.

Below the DSB strong interaction scale, there are only leptons in the fermion
sector,
and the lepton
number current is violated by an SU(2) anomaly which involves two
SU(2) currents. However, its
amplitude will vanish according to our results since the Higgs boson mass
in a composite Higgs theory is effectively infinite and hence $E^{spha}$
is also infinity.  Is this understandable? In the
fundamental lagrangian of quarks and leptons,
there are two kind of fermion number currents, one lepton number,
and the other baryon number. Both have an
SU(2) anomaly, however, their difference is anomaly free,
which means that the total change
of the lepton number must equal to the amount of baryon
 number change.
Since baryon fields do not exist in the low energy lagrangian[F.3]
lepton number violation process is forbidden. In other words, $E^{spha}$
should be infinity.

Another example which is also an application of our result
is the one-family technicolor model which has an electroweak sector
similar to that of the ordinary, light fermions[8].
In this model, both the techni quarks and the ordinary quarks have
an SU(2) anomaly. We can arrange the quantum
numbers of the techni quarks such that the sum of the two baryon
numbers is anomaly free.
Below the technicolor scale, the physics is described by an effective
theory
where techni fermions are integrated out. The resultant effective
lagrangian
is similar, in part, to the meson effective lagrangian of QCD[9].
Therefore, higher dimension operators
similar to ${\tilde {\cal O}}$ again appear.  As it is argued above,
$E^{spha}$ should be infinity[F.4]. Therefore, in the scenario of such a
model, baryon number violation at the electroweak scale is
forbidden as the sphaleron energy is infinity. This recovers the fact
that the overall baryon number, the sum of the techni and ordinary quarks,
is conserved, even though only the ordinary
fermions are the active degree of freedom in the low energy regime.

In conclusion, we took the standard model Higgs sector as an effective
theory with higher dimension operators added, and calculated the correction
of these higher dimension operators to the sphaleron energy. We found that
the sphaleron energy diverges in the large Higgs boson mass limit.
Without the higher dimension operators the sphaleron energy is stable
against the variation of the Higgs boson mass.
We have argued that there must be new physics above the Landau pole location
and shown that the operator making sphaleron energy diverge does exist
for example, in dynamical symmetry breaking models.
And in those models, our result becomes obvious.

\vfill\eject
X.Z. would like to thank R.N. Mohapatra and Y. Shen for discussions.
This work is performed in part at the Ames Laboratory under the Contract
No. W-7404-Eng-82 with the U.S. Department
of Energy Grants Nos. DE-FG-02-85ER40214
and
DE-FG02-87ER70371, Division of High Energy and Nuclear Physics.

\b
\ce {\bf Footnotes}
\b
\item{[F.1]}{The inclusion of the $U(1)_Y$, with the experimental
value of the Weinberg angle, has very little
effect on the sphaleron energy[3].}

\item{[F.2]}{For simplicity,
as usually we consider the limit of vanishing mixing angle $\Theta_W$,
so the $U(1)_Y$ field decouples.}

\item{[F.3]}{One would expect solitons to
 exist at the vacuum expectation value
scale which carry the baryon number[7]. However, these solitons could not
be produced energetically at low energy. Furthermore, the soliton solutions
are irrelevant to our discussions in this paper.}

\item{[F.4]}{Integrating out a generation of heavy fermion,
where the heavy fermion masses arise from Yukawa couplings,
will give expressions such as ${[\partial_\mu \phi \partial^\mu \phi
]}^2 / {\phi}^4$[10, 11]. This term makes $E^{spha}$ diverge for any
value of the Higgs boson
mass
because of the $\phi^4$ in the
${\it denominator}$, and the effective lagrangian is non-analytic.
Recently, H. Georgi, L. Kaplan and D. Morin[11] pointed out that
one must truncate the expansion of the
lagrangian in powers of the ${\it shifted}$
Physical Higgs field
to get an well-defined effective lagrangian
 and argued that the instanton action in the effective
theory is infinity due to the existence of a operator similar to
 $\tilde{\cal O}$. The operator ${\cal O}$ in this paper is analytic
and can make $E^{spha}$ diverge only for infinity Higgs boson mass.}

\vfill\eject

\ce{\bf References}
\b
\item{[1]}N.S. Manton, Phys. Rev. D28, 2019 (1983);
         F.R. Klinkhamer and N.S. Manton, Phys. Rev. D30, 2212 (1984).

\item{[2]}
      T. Akiba, H. Kikuchi and
T. Yanagida, Phys. Rev. D38, 1937 (1988); L.G. Yaffe, Phys. Rev. D40, 3463
 (1989); J. Kunz and Y. Brihaye, Phys. Lett. B216, 353 (1989).

\item{[3]}F. Klinkhamer and R. Lateveer, Z. Phys. C53, 247 (1992);
       B. Kleihaus, J. Kunz and Y. Brihaye, Phys. Lett. B273, 100 (1991);
      M.E.R. James, Z. Phys. C55, 513 (1992).

\item{[4]}B. Kastening, R.D. Peccei and X. Zhang, Phys. Lett. B266, 413 (1991);
 B. Kastening and X. Zhang, Phys. Rev. D45, 3884 (1992);
         K. Enqvist and I. Vilja, Phys. Lett. B287, 119 (1992);
S. Lee, J. Spence and B.-L. Young, Iowa State University Preprint,
 IS-J 5110, June 1993.

\item{[5]}S. Cortese, E. Pallante and R. Petronzio,
            Phys. Lett. B301, 203 (1993).

\item{[6]}E. D'Hoker and E. Farhi, Nucl. Phys. B248, 59 (1984);
            P.O. Simic, Phys. Rev. D36, 34 (1986)

\item{[7]}E. D'Hoker and E. Farhi, Phys. Lett. B134, 86 (1984);
           Nucl. Phys. B241, 109 (1984).

\item{[8]}E. Farhi and L. Susskind, Phys. Rept. 74, 277 (1980).

\item{[9]}R. Johnson,
            B.-L. Young, and D.W. McKay, Phys. Rev. D42, 3855 (1991).

\item{[10]}T. Banks and A. Dabholkar, Phys. Rev. D46, 4016 (1992).

\item{[11]}H. Georgi, L. Kaplan and D. Morin, Howard University Preprint,
                 HUTP-93/A030, 10/93.

\bye

\bye